\documentclass[12pt]{iopart}
\usepackage{epsf}
\begin{document}
\title{Experiments on the magnetorotational instability
in helical magnetic fields}

\author{Frank Stefani$^1$, Thomas Gundrum$^1$, Gunter Gerbeth$^1$, 
G\"unther R\"udiger$^2$, Jacek Szklarski$^2$, and Rainer Hollerbach$^3$}

\address{$^1$ Forschungszentrum Dresden-Rossendorf,
P.O. Box 510119, D-01314 Dresden, Germany}
\address{$^2$ Astrophysikalisches Institut Potsdam,
An der Sternwarte 16, D-14482 Potsdam, Germany}
\address{$^3$ Department of Applied Mathematics,
University of Leeds, Leeds, LS2 9JT, United Kingdom}

\begin{abstract}
The magnetorotational instability (MRI) plays a key role in the formation
of stars and black holes, by enabling outward angular momentum transport in
accretion disks. The use of combined axial and azimuthal magnetic fields
allows the investigation of this effect in liquid metal flows at moderate
Reynolds and Hartmann numbers. A variety of experimental results is
presented showing evidence for the occurrence of the MRI in a
Taylor-Couette flow using the liquid metal alloy GaInSn.
\end{abstract}
\pacs{47.20.-k, 47.65.+a, 95.30.Qd}
\submitto{\NJP}
\maketitle

\section{Introduction \label{intro}}
The existence of compact astrophysical objects relies on the fact that
they accumulate matter from their surroundings. Typically, the matter
around stars and black holes has organized itself into so-called accretion
disks. Before falling into the central object the rotating gas must be
slowed down. Unlike the energy, which can be radiated away, the angular
momentum of the rotating gas can only be transported within the disk.
The molecular viscosity of the gas is much too small to explain the
angular momentum transport needed to account for the accretion rates of
stars and black holes. Turbulent flows would suffice, but the onset of
turbulence would appear to be forbidden by the Rayleigh criterion,
stating that differentially rotating fluids become unstable only if their 
angular momentum decreases outward \cite{RAYLEIGH}. This obviously does
not hold true for Keplerian flows, whose angular momentum increases as
the square root of the radius.  It is widely believed \cite{RIZA} that
finite amplitude instabilities would ultimately destabilize even such
linearly stable flows, but recent experiments on rotating flows, with
carefully controlled axial boundary conditions, seem to indicate the
opposite \cite{JINATURE}.

This intriguing discussion about the possible role of finite amplitude
instabilities was circumvented by Balbus and Hawley in a seminal 1991
paper \cite{BAHA}. They showed that even weak magnetic fields dramatically
alter the stability criterion of rotating flows.  In fact, the basic idea
of this ``magnetorotational instability'' (MRI) was not completely new.
As early as 1959, Velikhov had demonstrated that an axial magnetic field
could destabilize a Rayleigh-stable Taylor-Couette (TC) flow, provided the
angular velocity decreases with radius \cite{VELI}. This result was later
confirmed by Chandrasekhar \cite{CHANDRA}, so the MRI may also be referred
to as the ``Velikhov-Chandrasekhar instability.''

For a TC flow driven by differentially rotating inner and outer cylinders,
the relevant stability boundaries are sketched in Fig.\ 1. If the ratio
$\mu:=\Omega_o/\Omega_i$ of the outer and inner cylinders' rotation rates
is less than the squared ratio $(r_i/r_o)^2$ of the inner and outer
cylinders' radii, then according to the Rayleigh criterion the flow is
always unstable (at least in the inviscid limit). In contrast, if $\mu$
is greater than one, then according to the Velikhov-Chandrasekhar
criterion the flow is stable.  The MRI occurs in the parameter regime
between these two lines, where the flow is hydrodynamically stable, but 
magnetohydrodynamically unstable.  And returning to the astrophysical 
application that Balbus and Hawley had in mind, we note that Keplerian
flows are precisely in this regime.

\begin{figure}[ht]
\begin{center}
\epsfxsize=10cm\epsfbox{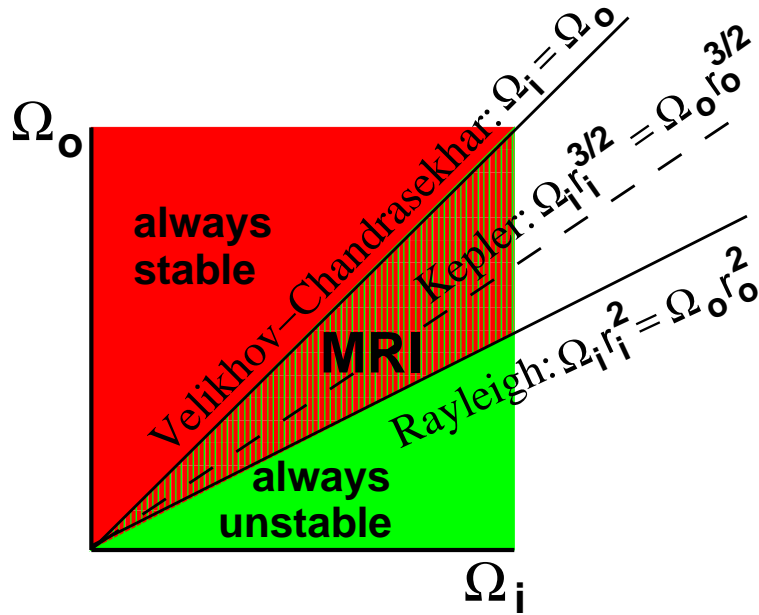}
\vspace{2mm}
\caption{Stability diagram of a Taylor-Couette flow with and without
magnetic fields. Whereas the hydrodynamic Rayleigh line separates
flows with increasing and decreasing {\it angular momentum}, the
magnetohydrodynamic Velikhov-Chandrasekhar line separates flows with
increasing and decreasing {\it angular velocity}. The dashed line
indicates a Keplerian profile, having angular velocity $\Omega\sim
r^{-3/2}$ decreasing, but angular momentum $\Omega r^2\sim r^{1/2}$
increasing.  (This particular example is for $r_i/r_o=\sqrt{1/2}$.)}
\end{center}
\end{figure}

As a result of this astrophysical importance of the MRI, there is
considerable interest in achieving it in laboratory experiments as well
\cite{ROSNER}.  For the classical Velikhov-Chandrasekhar (and also
Balbus-Hawley) configuration, with the externally applied magnetic field
being purely axial, these attempts have not entirely succeeded so far.
The underlying reason for this is that an azimuthal field, which is
necessary for the MRI to proceed, must then be produced from the applied
axial field by the rotation of the flow. This is only possible in flows
with sufficiently large magnetic Reynolds numbers $Rm$ (defined as the
product of magnetic permeability, electrical conductivity, length scale
and mean velocity of the flow). Such large magnetic Reynolds numbers are
very difficult to achieve, typically only in sodium-cooled fast-breeder
reactors or in special dynamo experiments \cite{RMP}. The only MRI
experiment to achieve such large values of $Rm$ is that of Lathrop's
group \cite{LATHROP}, who obtained an instability whose dependence on
$Rm$, as well as on the field strength, is indeed rather similar, if
perhaps not identical to, the classical MRI. However, this instability
arose from an already highly turbulent background flow, contradicting
the original goal of identifying the MRI as the {\it first} instability
on a laminar background flow.

Given that it is so difficult to produce the required azimuthal field
by the flow, why not simply replace the induction process by externally
applying an azimuthal magnetic field as well? This question was addressed
in a recent paper by Hollerbach and R\"udiger \cite{HORU}, who showed that
the MRI is then possible with far smaller Reynolds ($Re$) and Hartmann
($Ha$) numbers. This new type of MRI, sometimes called the ``helical MRI''
\cite{LIU}  or the ``inductionless MRI'' \cite{PRIEDE}, is currently the
subject of intense discussions in the literature \cite{LIU,PRIEDE,AN,
ANJACEK,BONANNO}. Indeed, the astrophysical relevance of magnetorotational
instabilities in helical magnetic fields is still a matter of some
controversy, dating back to an early dispute between Knobloch
\cite{KNOBLOCH1,KNOBLOCH2} and Hawley and Balbus \cite{BAHA2}.

Notwithstanding this ongoing discussion, the dramatic decrease of the
critical Reynolds number for the onset of the MRI in helical magnetic
fields, as compared with the case of a purely axial field, makes this new
type of MRI very attractive for experimental studies. Initial results of
the experiment ``PROMISE'' ({\it P\/}otsdam {\it RO\/}ssendorf
{\it M\/}agnetic {\it I\/}n{\it S\/}tability {\it E\/}xperiment)
were published recently \cite{PRL,APJL}. In this paper we report further
results. In particular, we document how the classical Taylor vortex flow
for $\mu=0$ changes to a very slowly traveling wave under the influence
of helical magnetic fields.

\section{The Experimental Facility}

The PROMISE facility, shown in Fig.\ 2, is a cylindrical Taylor-Couette
cell with externally imposed axial and azimuthal (i.e., helical) magnetic
fields. Its primary component is a cylindrical copper vessel V, fixed on
a precision turntable T via an aluminum spacer D.  The inner wall of this
vessel is 10 mm thick, extending in radius from 22 to 32 mm; the outer
wall is 15 mm thick, extending from 80 to 95 mm.  The outer wall of this
vessel forms the outer cylinder of the TC cell.  The inner cylinder I, also
made of copper, is fixed on an upper turntable, and is then immersed into
the liquid metal from above. It is 4 mm thick, extending in radius from
36 to 40 mm, leaving a 4 mm gap between it and the inner wall of the
containment vessel V. The actual TC cell therefore extends in radius from
40 to 80 mm, for a gap width $d=r_o-r_i=40$ mm.  The fluid is filled to a
height of 410 mm, for an aspect ratio of $\sim\!10$.

\begin{figure}[ht]
\begin{center}
\epsfxsize=14cm\epsfbox{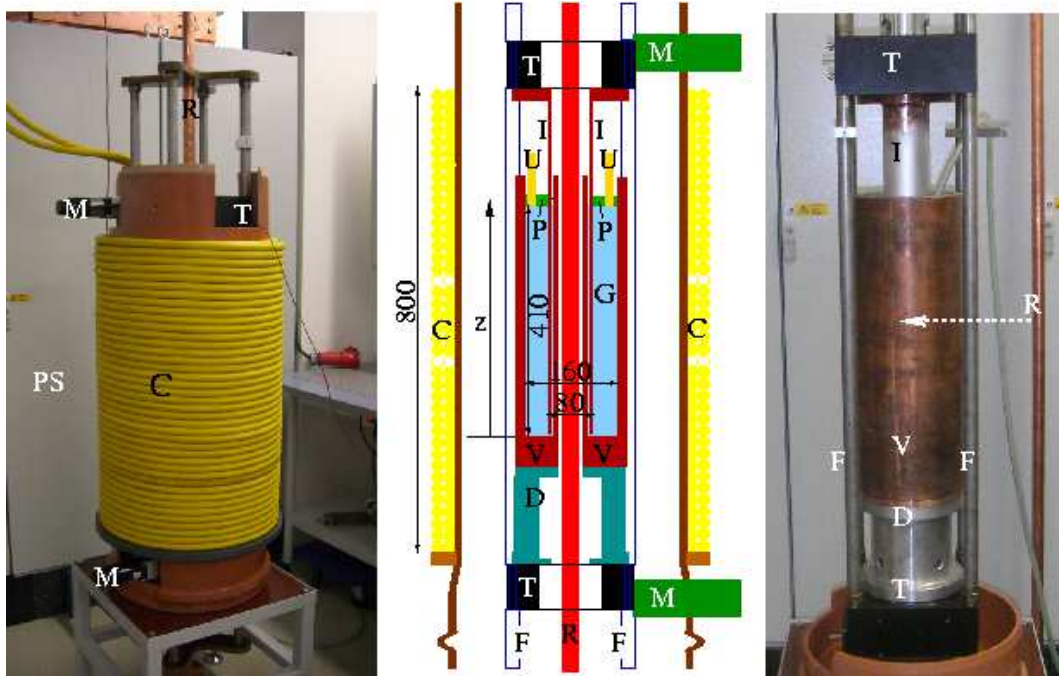}
\vspace{2mm}
\caption{The left panel shows the complete PROMISE facility; the
right panel shows the central module, without the coils C, and with
the rod R taken out of the center.  The middle panel shows a
schematic sketch, with the numbers indicating lengths in mm, and the
letters denoting the following:
V - Copper vessel, I - Inner cylinder, G - GaInSn,
U - Two ultrasonic transducers,
P - Plexiglass lid,
T - High precision turntables,
M - Motors,
F - Frame,
C - Coil,
R - Copper rod,
PS - Power supply up to 8000 A.}
\end{center}
\end{figure}

The fluid is the eutectic alloy Ga$^{67}$In$^{20.5}$Sn$^{12.5}$, which
is liquid at room temperatures. The physical properties of GaInSn at 25
$^{\circ}$C are as follows: density $\rho=6.36 \times 10^3$ kg/m$^3$,
kinematic viscosity $\nu=3.40\times 10^{-7}$ m$^2$/s, electrical
conductivity $\sigma=3.27\times 10^6$ ($\Omega$ m)$^{-1}$. The magnetic
Prandtl number is $Pm=\mu_0\sigma\nu=1.40 \times 10^{-6}$.

In the current experimental setup, the upper endplate of the TC cell
is a plexiglass lid P, fixed to the frame F, and hence stationary.
The lower endplate is simply part of the copper vessel V, and thus
rotates with the outer cylinder.  There is therefore a clear asymmetry
in the endplates, with regard to both their rotation rates and
electrical conductivities.

Hydrodynamic TC experiments are typically done using glass cylinders,
which allow for very good geometrical accuracy, to within $\sim\!10^{-2}$
mm \cite{SCHULTZGRUNOW}. We used copper cylinders here because the
critical Reynolds and Hartmann numbers for the onset of the MRI are
somewhat smaller with perfectly conducting boundaries than with
insulating boundaries \cite{AN}, suggesting that copper would be more
suitable than glass or plexiglass. However, the price to pay for this
is that $\sim\!10^{-2}$ mm accuracy is no longer achievable when drilling
and polishing a material as soft as copper. What is more, to ensure a
well-defined electrical contact between the fluid and the walls, it is
necessary to intensively rub the GaInSn into the copper; the resulting
abrasion then limits the accuracy to no better than $\sim\!10^{-1}$ mm.

The rotation frequencies of the inner and outer cylinders are measured by
the Reynolds number $Re=2 \pi f_i r_i d/\nu$ and the ratio $\mu=f_o/f_i$.
Typical Reynolds numbers in the experiment are $O(10^3)$, some 30
times greater than the $Re_c=68.2$ onset of nonmagnetic Taylor vortices
in TC flows with $\mu=0$ (for a radius ratio $\eta:=r_i/r_o$ of 0.5).
For the rotation ratio $\mu$, we will consider both $\mu<0.25$, for
which the basic flow profile is already Rayleigh-unstable, as well as
$\mu>0.25$, for which it is Rayleigh-stable, but MRI-unstable.  One of
the issues then that we will particularly want to focus on is how the
behavior changes as we cross the Rayleigh line $\mu=0.25$.

Axial magnetic fields of order 10 mT are produced by a double-layer coil
(C). Windings were omitted at two symmetric positions close to the middle
in order to optimize the homogeneity of the field throughout the region
occupied by the fluid. Currents up to 200 A are driven through this coil,
achieving axial fields up to $B_z=20.35$ mT, or in nondimensional units
up to a Hartmann number $Ha:=B_z (r_i d \sigma/\rho \nu)^{1/2}$ of 31.65.
The azimuthal field $B_\varphi$, also of order 10 mT (at $r_i$), is
generated by a current through a water-cooled copper rod (R) of radius 15
mm. The power supply for this axial current delivers up to 8000 A. In the
following this current will be referred to as the ``rod current.''

In nonmagnetic TC experiments the flow can be visualized and measured by
a wide variety of techniques.  In contrast, making measurements in liquid
metal flows is non-trivial.  The first measurements of axial velocities in
liquid metal TC flows were by Takeda \cite{TAKEDA}, using Ultrasound Doppler
Velocimetry (UDV).
Applying a similar technique \cite{CRAMER}, our 
measuring instrumentation consists simply
of two ultrasonic transducers, fixed into the upper plexiglass lid, 15 mm
away from the outer copper wall, flush mounted at the interface to the
GaInSn, and with special high-focus sensors having a working frequency of
4 MHz.  With this instrumentation we can then measure the axial velocity
$v_z$ at the particular location $r=65$ mm (averaged over the
approximately 8 mm width of the ultrasound beam), as a function of time
$t$ and height $z$ along the cylinder axis.  The resolutions in $t$ and
$z$ were adjustable (within limits); in most runs we used resolutions of
1.84 sec in $t$ and 0.685 mm in $z$.  Finally, having two transducers, on
opposite sides of the TC cell, was important in order to be able to
distinguish between the expected axisymmetric ($m=0$) MRI \cite{HORU,AN},
and an unexpected non-axisymmetric $m=1$ instability which arose in
certain parameter regimes.

\section{The Dependence on $\mu$}

\subsection{Nonmagnetic results}
Some preliminary experiments on the onset of nonmagnetic Taylor vortex
flow (TVF) were done with a water-glycerol mixture.  Thereafter we
filled the vessel with the GaInSn, but continued to do nonmagnetic runs.
Measuring the initial onset of TVF was unfortunately not possible,
because the velocities corresponding to $Re=68$ were too small to
measure with our ultrasound technique. (The water-glycerol mixture
had a considerably greater viscosity, so $Re=68$ translates into
correspondingly greater velocities there, making those measurements
easier, and a useful test of the basic ultrasound technique.)

Even though obtaining the initial onset of the classical $\mu=0$ TVF
was not possible for the GaInSn, we were able to measure how the
existing, super-critical TVF gradually disappears again as we increase
$\mu$ beyond the Rayleigh line $\mu=0.25$. Figure 3 shows measurements
for eight different values of $\mu$, from 0 to 0.3. The inner cylinder's
frequency was fixed at $f_i=0.1$ Hz, corresponding to $Re=2958$; the
outer cylinder's frequency was then adjusted to yield the indicated
values of $\mu$. For $\mu=0$, the alternating red and blue stripes
indicate a steady TVF with 5 pairs of rolls along the vertical axis,
exactly as we would expect for a TV cell of aspect ratio 10. Increasing
$\mu$, this TVF gradually breaks up in both space and time, eventually
disappearing completely. For $\mu=0.25$ and 0.3 there is no trace of
TVF; what we see instead is simply an Ekman flow driven by the top
and bottom endplates. We recall that the upper lid is stationary, whereas
the lower lid rotates with the outer cylinder. In both cases this drives
a radially inward Ekman pumping, leading to $v_z$ being positive/negative
in the upper/lower halves of the cell \cite{CZARNY}. Note also how these
two Ekman vortices are asymmetric about the midplane, due to the stronger
pumping at the upper, stationary lid. The two regions are separated by a
boundary containing a jet-like radial outflow, as discussed in \cite{KAGE}.

\begin{figure}[ht]
\begin{center}
\epsfxsize=12cm\epsfbox{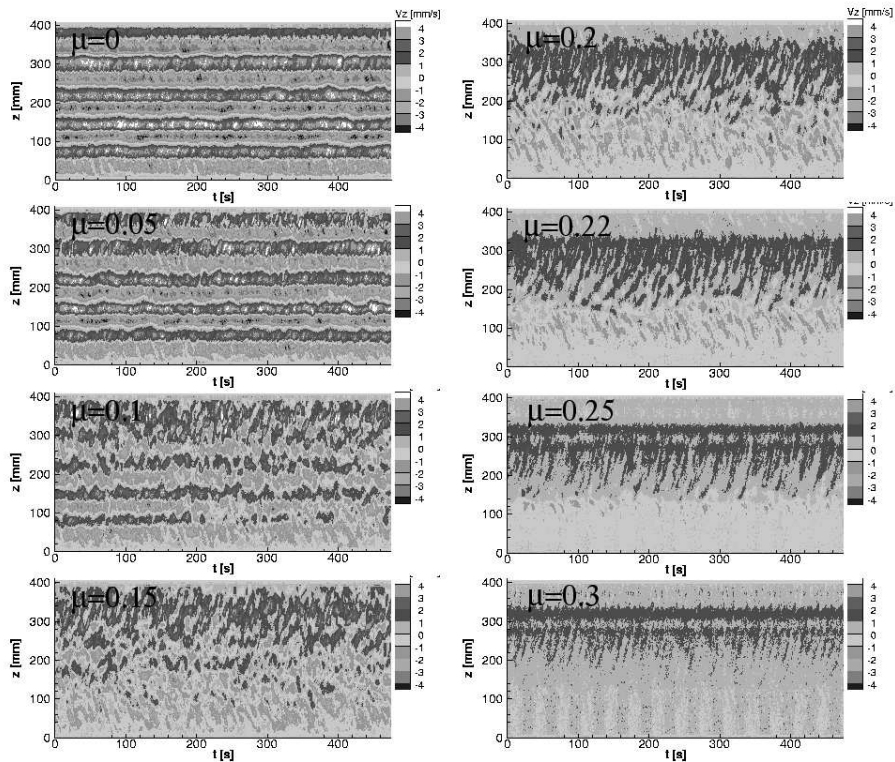}
\vspace{2mm}
\caption{$v_z$ at $r=65$ mm, as a function of time $t$ and height $z$.
The inner cylinder's rotation rate is fixed at $f_i=0.1$ Hz ($Re=2958$);
the outer cylinder's rotation rate is adjusted to yield the indicated
values of $\mu=f_o/f_i$. $I_{\rm coil}=I_{\rm rod}=0$, so these results
are completely nonmagnetic.}
\end{center}
\end{figure}

\subsection{$\mu=0$, purely axial or purely azimuthal fields}
Before considering the effect of helical magnetic fields, it is
interesting to examine the influence of purely axial or purely azimuthal
fields, neither of which yields an MRI at these small values of $Re$.
Figure 4 shows $v_z$ in these cases, namely a purely axial field on the
left, and a purely azimuthal field on the right, and $Re=2958$ and $\mu=0$
for both.  We note how the basic TVF structure is essentially the same as
in the nonmagnetic case, apart from minor changes in strength, and a
slight fluctuation in time. In subsequent sections we will see that
helical fields do indeed yield rather different results, just as
predicted theoretically.

\begin{figure}[ht]
\begin{center}
\epsfxsize=12cm\epsfbox{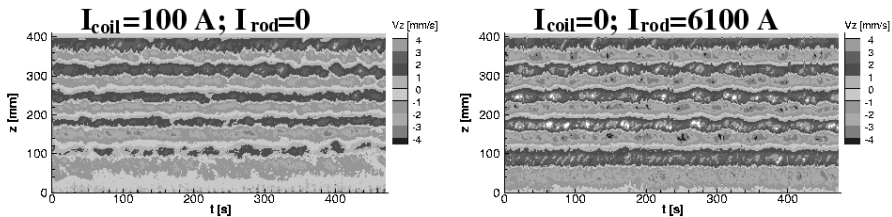}
\vspace{2mm}
\caption{The left panel has a purely axial field, with $I_{\rm coil}=100$
A; the right panel has a purely azimuthal field, with $I_{\rm rod}=6100$
A. The inner cylinder rotates at $f_i=0.1$ Hz ($Re=2958$); the outer
cylinder is stationary ($\mu=0$).}
\end{center}
\end{figure}

\subsection{Increasing $\mu$, helical magnetic fields}

Purely axial and purely azimuthal magnetic fields both have the property
that $\pm z$ are identical (apart from the asymmetries introduced by the
different boundary conditions on the top and bottom endplates).  This is
reflected in Figs.\ 3 and 4, which show relatively little top/bottom
asymmetry. In contrast, helical magnetic fields break this reflectional
symmetry, as first pointed out by Knobloch \cite{KNOBLOCH2}.  As a result
of this symmetry breaking, the previously steady TVF is forced to drift
in $z$, resulting in a traveling-wave TVF (still axisymmetric though).
The direction of propagation, that is, whether the pattern drifts in the
$+z$ or $-z$ directions, depends on whether the screw-sense of the
magnetic field is either parallel or anti-parallel to the flow rotation.
For a given helical field, the waves therefore propagate in one direction
only; standing waves do not arise in this problem.

Figure 5 shows results for the helical field generated by the currents
$I_{\rm coil}=60$ A and $I_{\rm rod}=6100$ A, corresponding to $Ha=9.5$,
and the ratio $\beta:=B_\varphi(r_i)/B_z=6.0$. The inner cylinder's
rotation is fixed at $f_i=0.05$ Hz, corresponding to $Re=1479$, and the
outer cylinder's rotation is again adjusted to yield the indicated
values of $\mu$. Comparing these results with the nonmagnetic results
in Fig.\ 3, we see that the two are already different for $\mu=0$.
We now have slightly inclined red and blue stripes, indicating that the
TVF rolls are no longer steady, but drift upward in time.  For $\mu=0$
the frequency of this wave is very small, around 1.6 mHz, only 3\% of
$f_i$. The frequency increases with increasing $\mu$, reaching
9 mHz, or 18\% of $f_i$.

We see therefore that the previously steady TVF has indeed been replaced
by a unidirectionally traveling-wave TVF. Furthermore, unlike in Fig.\ 3,
where the TVF rolls disappeared as we increased $\mu$ beyond the Rayleigh
line, in this case these TVF waves continue to exist at $\mu=0.25$ and
even 0.27, gradually fading away only for $\mu=0.3$ and 0.35. These
traveling-wave instabilities beyond the Rayleigh line are precisely the
theoretically predicted MRI.

\begin{figure}[ht]
\begin{center}
\epsfxsize=12cm\epsfbox{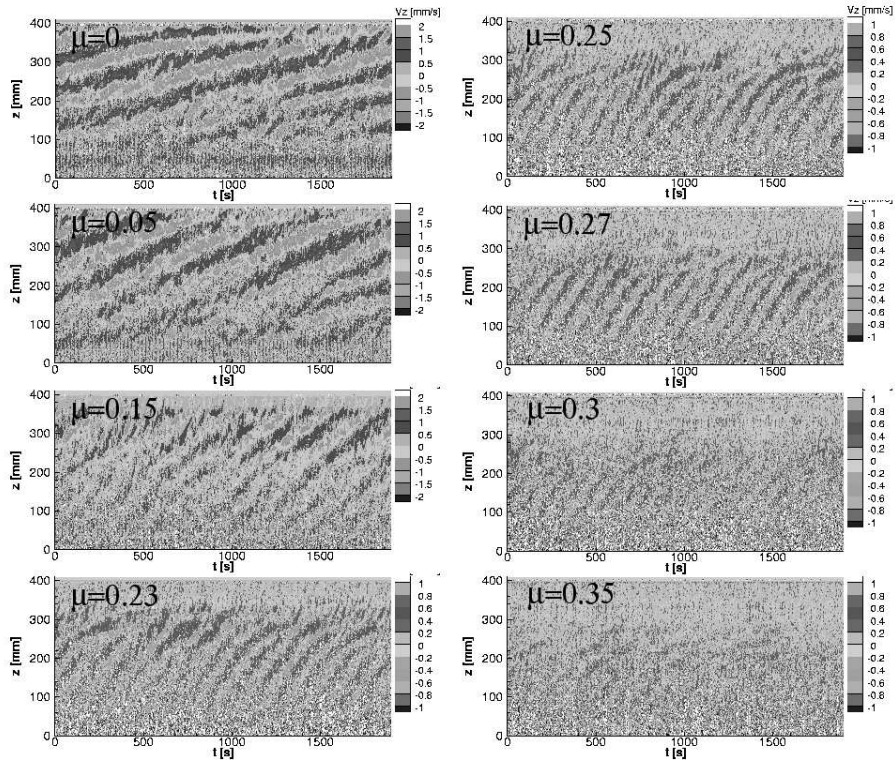}
\vspace{2mm}
\caption{$I_{\rm coil}=60$ A, corresponding to $Ha=9.5$; $I_{\rm rod}=
6100$ A, corresponding to $\beta:=B_\varphi(r_i)/B_z=6.0$. The inner
cylinder's rotation rate is fixed at $f_i=0.05$ Hz ($Re=1479$); the outer
cylinder's is adjusted according to the indicated values of $\mu$. Note
how the frequency of the waves increases with increasing $\mu$.}
\end{center}
\end{figure}

It is interesting to compare the measured wave frequencies with
predictions from two different numerical methods. First, we can solve the
1-dimensional linear eigenvalue problem in an axially unbounded cylinder,
with perfectly conducting inner and outer boundaries \cite{HORU,SHALYBKOV,
AIP}. The second method is to solve the full 2-dimensional nonlinear
problem in a properly bounded cylinder, as presented in \cite{ANJACEK}.

The left panel in Fig.\ 6 shows how the experimental Reynolds number
$Re=1479$ compares with the critical Reynolds number obtained from the
1D linear eigenvalue analysis; we see that at $\mu=0$ the experiment is
some 7 times supercritical, but at $\mu=0.27$ it is only about twice
supercritical, and at $\mu=0.3$ it is slightly subcritical, in agreement
with Fig.\ 5, where the wave indeed fades away between $\mu=0.27$ and 0.3.
The right panel in Fig.\ 6 compares the measured and computed wave
frequencies, normalized to $f_i$. The overall agreement 
is quite good. Note how both the
measured values and the 2D simulation yield a slightly slower increase
with $\mu$ than that predicted by the (more highly idealized) 1D analysis.

\begin{figure}[ht]
\begin{center}
\epsfxsize=12cm\epsfbox{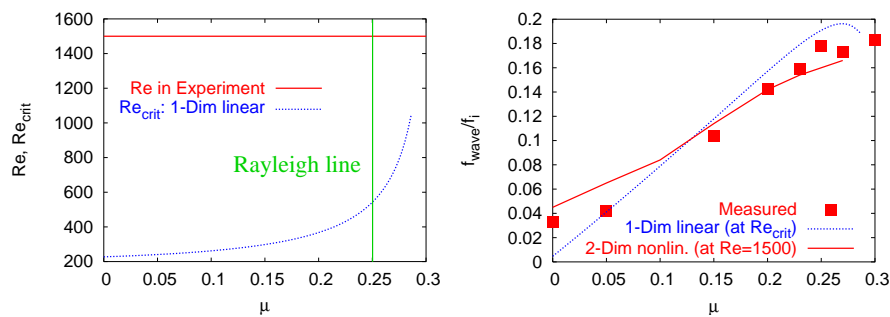}
\vspace{2mm}
\caption{The left panel compares the experimental Reynolds number
$Re=1479$ with the critical value from the 1D eigenvalue problem. The
right panel compares the experimental wave frequencies with results from
the 1D and 2D numerical calculations.}
\end{center}
\end{figure}

Finally, Fig.\ 7 shows the detailed spatio-temporal structure of the
$\mu=0.27$ 2D simulation, and compares it with the corresponding panel
from Fig.\ 5. The magnitudes of $v_z$ are not quite the same, but
otherwise the agreement is rather good.  It is particularly gratifying
to note that the waves are concentrated between $z\approx50$ and 250 in
both the experiment and the simulation, suggesting that the simulation is
properly reproducing the end-effects that cause the waves to fade away at
the ends. The frequencies and wavelengths are also in reasonable agreement
between the two panels.

\begin{figure}[ht]
\begin{center}
\epsfxsize=12cm\epsfbox{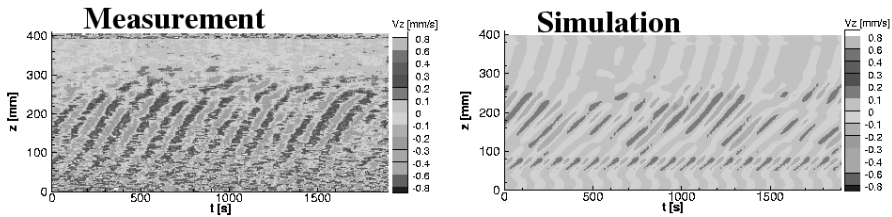}
\vspace{2mm}
\caption{The left panel is the same as the $\mu=0.27$ panel in Fig.\ 5.,
but with a moving average over 20 data points and a modified velocity 
scale.
The right panel is from the corresponding 2D numerical simulation, also
at $\mu=0.27$, $Re=1479$, $Ha=9.5$ and $\beta=6.0$.}
\end{center}
\end{figure}

\section{Increasing $B_z$, for fixed $B_\varphi$}

In the previous section we have considered the dependence on $\mu$, for
various magnetic field configurations. We now fix $\mu=0.27$ ($f_i=0.06$
Hz, $f_o=0.0162$ Hz, $Re=1775$), and consider the dependence on the field.
We begin with Fig.\ 8, showing the effect of a purely azimuthal field (so
similar to Fig.\ 4b, but now beyond the Rayleigh line, rather than at
$\mu=0$). Without a field (the left panel), the flow does not exhibit any
coherent structures at all, certainly nothing like TVF cells. There is a
slight modulation at the frequency $f_o$, probably due to geometrical
imperfections of the outer cylinder, or perhaps oxides sticking to it.
Switching on the azimuthal field (the right panel), these structures
become somewhat more pronounced, but continue to be restricted to a very
narrow region in the center of the apparatus. There is certainly nothing
like the traveling-wave structures observed in Fig.\ 5.  This is of
course only to be expected, since a purely azimuthal field does not yield
an MRI, at least not at such a small value of $Re$.

What we wish to consider next is what happens if we now gradually switch
on the axial field. Once it is sufficiently strong, we would certainly
expect to recover the MRI, just as in Fig.\ 5. If we continue increasing
the axial field though, we would also expect the MRI to disappear again,
since it only exists within a certain range of field strengths, but
disappears if the field is either too weak or too strong. This well-known
feature of the MRI was previously documented in \cite{PRL}; here we
substantiate it in more detail.

Figure 9 therefore shows a sequence of eight runs, all at the previous
values $\mu=0.27$, $Re=1775$, $I_{\rm rod}=6000$ A, and $I_{\rm coil}$
increasing from 20 to 140 A as indicated.  At 20 A we still have a rather
featureless flow, similar to Fig.\ 8 (in fact, more like the nonmagnetic
left panel than the magnetic right panel). However, at 40 A we already
see the same traveling-wave MRI as in Fig.\ 5. Increasing $I_{\rm coil}$
further, this mode increasingly fills the apparatus, until at 80 and 90 A
it extends over almost the entire height. Finally, at 100 A this mode
begins to disappear again, and by 140 A we are back to the same
featureless flow we started out with.

\begin{figure}[ht]
\begin{center}
\epsfxsize=12cm\epsfbox{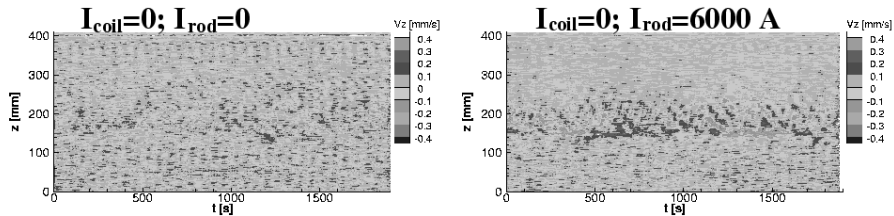}
\vspace{2mm}
\caption{$\mu=0.27$, $Re=1775$ and $I_{\rm coil}=0$ for both panels; the
left panel also has $I_{rm rod}=0$, the right panel has $I_{\rm rod}=6000$
A. Note the absence of any TVF structures.}
\end{center}
\end{figure}

\begin{figure}[ht]
\begin{center}
\epsfxsize=12cm\epsfbox{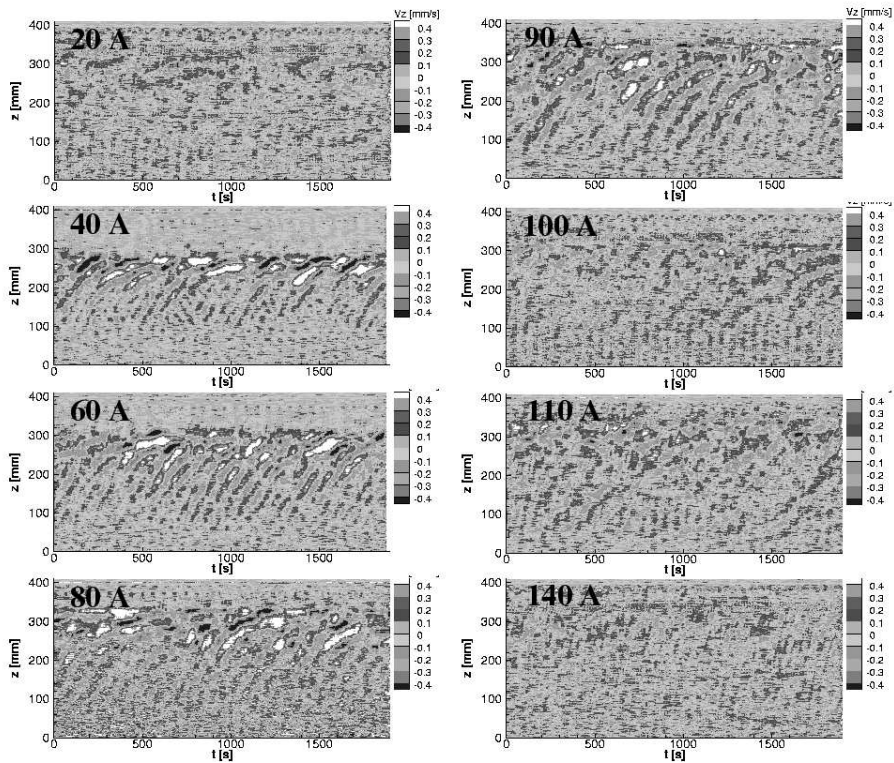}
\vspace{2mm}
\caption{$v_z$ as a function of $t$ and $z$, with the rotation rates
fixed at $\mu=0.27$ and $Re=1775$, the azimuthal field fixed at
$I_{\rm rod}=6000$ A, and the axial field increasing from $I_{\rm coil}
=20$ to 140 A as indicated. The traveling-wave MRI only exists for the
intermediate range $I_{\rm coil}=40$ to 100 A.}
\end{center}
\end{figure}

Figures 10 and 11 show further details of the behavior at 70 A, so right
in the middle of the MRI regime in Fig.\ 9. In particular, Fig.\ 10
compares the signals from the two ultrasound transducers, and shows that
these modes are indeed the same at both sensors, as they ought to be for
an axisymmetric instability mode. Figure 11 also shows both sensors, but
now shows the effect of suddenly switching on the field (both $B_z$ and
$B_\varphi$ simultaneously). Ideally one would like to use this data to
extract the linear growth rate of the instability, and compare it with the
theoretical predictions. The UDV technique unfortunately cannot deliver very
small values of $v_z$ accurately enough to obtain a definite growth rate,
but we can certainly observe that the MRI is fully developed 200--300 sec
after switching on the field, consistent with the expected growth rate
of $\sim\!0.1f_i$.

\begin{figure}[ht]
\begin{center}
\epsfxsize=12cm\epsfbox{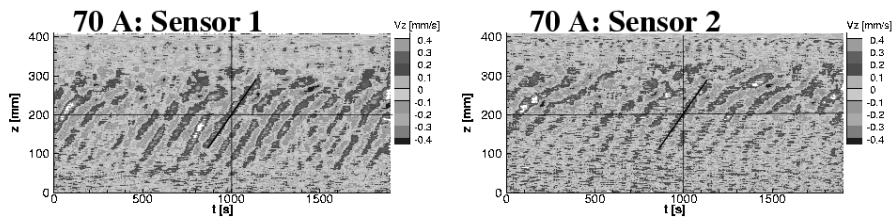}
\vspace{2mm}
\caption{Measurements made by the two ultrasound sensors. The inclined
bars show that the two signals are in phase, as required for an
axisymmetric mode. $\mu=0.27$, $Re=1775$, $I_{\rm rod}=6000$ A,
$I_{\rm coil}=70$ A.}
\end{center}
\end{figure}

\begin{figure}[ht]
\begin{center}
\epsfxsize=12cm\epsfbox{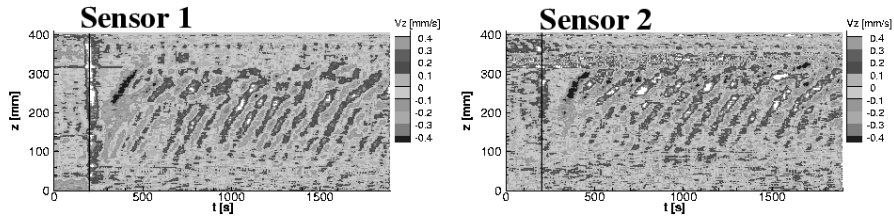}
\vspace{2mm}
\caption{The field ($I_{\rm rod}=6000$ A, $I_{\rm coil}=70$ A) is suddenly
switched on at $t=200$ sec, indicated by the black lines. Some 200--300 sec
later the traveling-wave MRI is essentially fully developed, consistent
with the expected growth rate of around $0.1f_i$. $\mu=0.27$, $Re=1775$.}
\end{center}
\end{figure}

Figure 12 compares the experimental results with some of the 1D
numerical results \cite{HORU,SHALYBKOV,AIP}. First, the red squares and
lines compare the measured frequencies with two different computed
frequencies, namely at $Re_{\rm crit}$, and at the experimental value
$Re=1775$. The difference between the two numerical results is fortunately
small, suggesting that the frequency in the nonlinear, saturated regime
(the regime the experiment is in) is most likely also close to these
values. The measured frequencies are certainly in reasonable agreement
with the computed ones, and in particular also exhibit a maximum around
$Ha=10$. 

Next, the dotted blue line is the numerically computed linear growth rate
at $Re=1775$, and shows that (a) at this value of $Re$ the MRI should exist
between $Ha=5$ and 17, or $I_{\rm coil}=30$ and 110 A, broadly consistent
with Fig.\ 9, and (b) the maximum growth rate, at $Ha=11$, is $0.1f_i$,
consistent with Fig.\ 11. Finally, the blue circles denote the measured rms
values of $v_z$. Note how these values are greatest in the same $5<Ha<17$
interval where we expect the MRI (although they are also significantly
different from zero outside this range).

\begin{figure}[ht]
\begin{center}
\epsfxsize=12cm\epsfbox{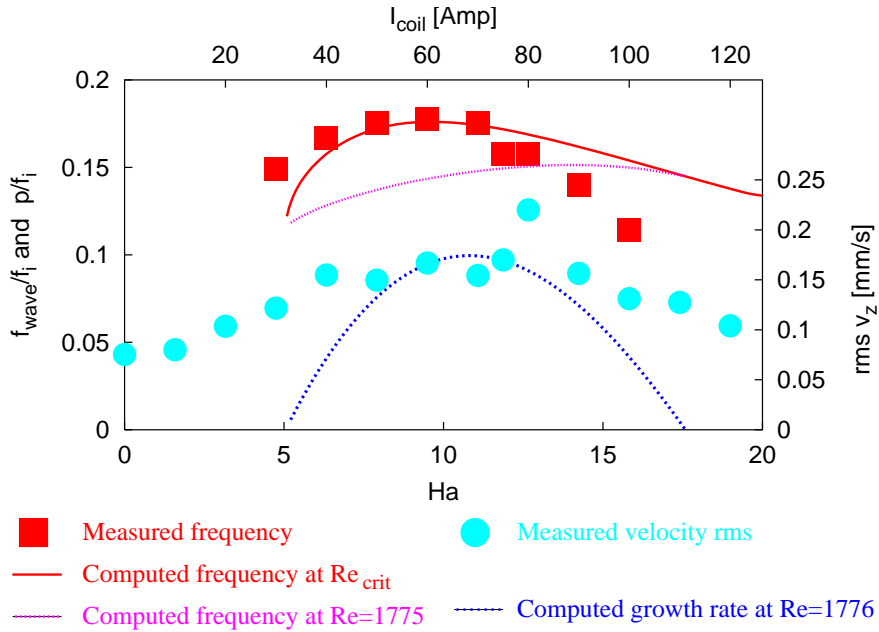}
\vspace{2mm}
\caption{Various comparisons of experimental (squares and circles) and
numerical (lines) results, as functions of $I_{\rm coil}$, or equivalently
$Ha$. $I_{\rm rod}=6000$ A, $\mu=0.27$ and $Re=1775$. Normalized frequencies
$f_{\rm wave}/f_i$ and growth rates $p/f_i$ are indicated on the left axis;
rms values of $v_z$ on the right axis.}
\end{center}
\end{figure}

\section{An Unexpected Non-axisymmetric Mode \label{nonaxi}}

All of the results presented so far are in general agreement with the
theoretical predictions for the axisymmetric traveling-wave MRI. In
contrast, Fig.\ 13 shows an unexpected non-axisymmetric mode, which
occurred for $f_i=0.05$ Hz, $f_o=0.0135$ Hz, $I_{\rm rod}=6720$ A,
and $I_{\rm coil}=110$ A, or in dimensionless numbers $\mu=0.27$,
$Re=1479$, $Ha=17.4$ and $\beta=3$. The two sensors are now out of phase,
indicative of a non-axisymmetric mode, most likely $m=1$. The frequency
of this mode is 0.21 Hz. Such a non-axisymmetric mode is not expected as
the first instability (although similar $m=1$ instabilities are known to
exist in the related Tayler instability problem \cite{NONAXISYMMETRIC}),
but could conceivably arise as a secondary instability of the axisymmetric
traveling-wave state, or could alternatively be due to a lack of perfect
axisymmetry in the apparatus (e.g. due to the field induced by the
$I_{\rm rod}$ current when it flows through the leads to the central rod).
Fully understanding this non-axisymmetric mode will require further work,
both experimental and numerical (i.e. extending the 2D nonlinear code to
3D).

\begin{figure}[ht]
\begin{center}
\epsfxsize=12cm\epsfbox{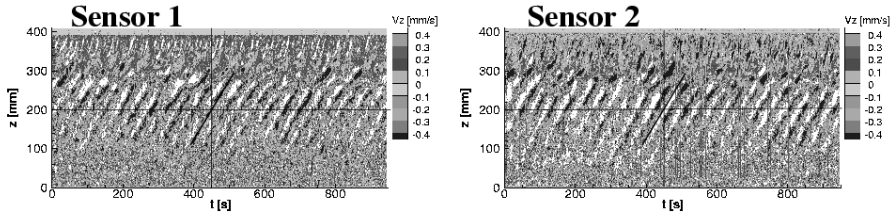}
\vspace{2mm}
\caption{$\mu=0.27$, $Re=1479$, $I_{\rm rod}=6720$ A, $I_{\rm coil}=110$
A. The inclined bars show that the two signals are out of phase,
corresponding to an $m=1$ mode.}
\end{center}
\end{figure}

\section{Conclusions and Future Prospects \label{conclusion}}

In this work we have presented experimental results that make a strong
case for the existence of a magnetorotational instability in the presence
of an externally applied helical magnetic field. In particular, we showed
that a unidirectionally traveling wave disturbance is excited, and
continues to exist beyond the Rayleigh line. The frequency of this wave,
as well as its dependence on the axial field strength, are in good
agreement with 1D and 2D numerical calculations. This traveling-wave MRI
differs in important ways from the classical MRI considered by Velikhov
\cite{VELI}, Chandrasekhar \cite{CHANDRA}, and Balbus and Hawley
\cite{BAHA}, but it does share the most basic features that any
magnetorotational instability ought to possess, namely that it would not
exist without the magnetic field, but on the other hand, is not driven by
the field, but instead draws all its energy from the differential rotation.
(The argument over the astrophysical relevance of helical magnetic fields
for the MRI \cite{KNOBLOCH1,KNOBLOCH2,BAHA2,LIU} will most likely continue
though.)

Regarding future experimental work, a modification of the apparatus is
currently under way, to symmetrize the axial boundary conditions by using
plexiglass at both ends. The use of split rings for the endplates, as
suggested by \cite{KAGE,HF}, and implemented in the (so far nonmagnetic)
experiment by \cite{JINATURE}, is also planned. This should significantly
reduce the Ekman circulation cells and the associated radial jet close to
the midplane. We hope that the transition to the MRI will then be
considerably sharper than it is in the results presented here.

There is also much numerical work that remains to be done, including a
more realistic treatment of the magnetic boundary conditions, both radial
and axial (copper is after all not a perfect conductor). Given the good
agreement obtained so far, this is unlikely to significantly affect the
results, but certainly needs to be investigated. Next, a 3D
finite-cylinder code needs to be developed, to study aspects like this
non-axisymmetric instability in section \ref{nonaxi}. Finally, from a
general theoretical point of view, there is an urgent need to clarify the
issue of convective versus absolute instabilities, which also arises in
many other problems related to TC flows \cite{ABSOLUT}.

\ack
This work was supported by the German Leibniz Gemeinschaft, within its SAW
program.  We thank Heiko Kunath for technical assistance, and Markus Meyer
for assistance taking some of the data.  The Rossendorf group also thanks
Janis Priede and Ilmars Grants for stimulating discussions.

\end{document}